\documentclass[12pt,preprint]{emulateapj}
\usepackage{natbib}
\usepackage{graphicx}
\usepackage{amssymb}
\usepackage{epstopdf}

\begin{document}

\shorttitle{CR excess due to reconnection}
\shortauthors{Lazarian \& Desiati}

\title{Magnetic reconnection as the cause of cosmic ray excess from the heliospheric tail}
\author{A. Lazarian}
\affil{Astronomy Department, University of Wisconsin, Madison, WI 53706 }
\author{P. Desiati}
\affil{Department of Physics, University of Wisconsin, Madison, WI 53706 }

\begin{abstract}
The observation of a broad excess of sub-TeV cosmic rays compatible with the direction of the heliospheric tail \citep{nagashima} and the discovery of two significant localized excess regions of multi-TeV cosmic rays by the MILAGRO collaboration \citep{abdo2}, also from the same region of the sky, have raised questions on their origin. In particular, the coincidence of the most significant localized region with the direction of the heliospheric tail and the small angular scale of the observed anisotropy ($\sim$ 10$^{\circ}$) is suggestive a local origin and of a possible connection to the low energy broad excess. Cosmic ray acceleration from magnetic reconnection in the magnetotail is proposed as a possible source of the energetic particles. 
\end{abstract}
\keywords{magnetic fields-- MHD-- solar wind--acceleration of particles--cosmic rays}

\section{Introduction}
\label{sec:intro}

It is known that cosmic rays arrival direction has an energy dependent large angular scale anisotropy with an amplitude of order $10^{-4}-10^{-3}$. The first comprehensive observation of this anisotropy was provided by a network of muon telescopes sensitive to sub-TeV energies and located at different latitudes \citep{nagashima}. More recently, an anisotropy was also observed in the multi-TeV energy range by the Tibet AS$\gamma$ array \citep{amenomori}, Super-Kamiokande \citep{guillian} and by MILAGRO \citep{abdo}, and the first high statistics observation in the southern hemisphere in the 10 TeV region, is being reported by IceCube \citep{abbasi}. The origin of the large angular scale anisotropy in the cosmic rays arrival direction is still unknown. The structure of the local interstellar magnetic field is likely to have an important role. However the combined study of the anisotropy energy and angular dependency, its time modulation and angular scale structure seem to suggest that the observation might be a combination of multiple superimposed effects, caused by phenomenologies at different distances from Earth.

In this context, particular interest is derived from the observation of a broad excess of sub-TeV cosmic rays in a portion of the sky compatible with the direction of the heliospheric tail (or heliotail) \citep{nagashima,hall} (see \S \ref{sec:obs}). The heliotail is the region of the heliosphere downstream the interstellar matter wind delimited within the heliopause, i.e. the boundary that separates the solar wind and interstellar plasmas \citep{izmodenov}. The observed excess was attributed to some unknown anisotropic process connected with the heliotail (thus called tail-in excess). The gyro-radius of sub-TeV cosmic protons is less than about 200 AU (in a $\sim$1 $\mu$G interstellar magnetic field), which is approximately the size of the heliosphere and, most likely, smaller than the width and length of the heliotail. The persistence of the cosmic ray anisotropy structure in the multi-TeV energy range makes it challenging to link this observation to the heliosphere. Although the unknown size and extension of the heliotail contributes to the uncertainty on the energy scale at which heliospheric influence on cosmic rays starts to be negligible. However, we know that the observations of multi-TeV cosmic rays anisotropy show small angular scale patterns superimposed to the smooth broad structure of the tail-in excess, which is suggestive of a local origin, i.e. within the heliotail. With the same technique used in gamma ray detection to estimate the background and search for sources of gamma rays, the MILAGRO collaboration discovered two localized excess regions in the cosmic rays arrival direction distribution \citep{abdo2}. The same excess regions were reported by the ARGO-YBJ air shower array \citep{vernetto}. The strongest and more localized of them (with an angular size of about 10$^{\circ}$) coincides with the direction of the heliotail. The peculiarity of such an observation triggered an astrophysical interpretation based on the possibility that cosmic rays accelerated by the supernova that produced Geminga pulsar are focussed by an ad-hoc interstellar magnetic field structure \citep{salvati,drury} (see \S \ref{sec:interp}).

The localized regions lie in the same portion of the sky that is dominated by the broad tail-in excess at lower energy, and, although it might be coincidental, we interpret this as manifestations of the same phenomenology at different energies.


It is proposed that both sub-TeV tail-in excess and the multi-TeV localized excess of cosmic rays might be caused by magnetic reconnection in the heliosphere and, in particular in the heliotail, where the distance scale might be long enough to induce sufficient acceleration at high energy. The very idea of appealing to magnetic reconnection for the acceleration of energetic particles can be traced back to pioneeding works by Giovanelli (1946) and Dungey (1953). The uncertainties with understanding of fast reconnection were one of the impediments for applying the process to energetic particle acceleration (see Lazarian \& Opher 2009). We appeal to the model of reconnection of weakly stochastic field in Lazarian \& Vishniac (1999), which was identified as a cause of First Order Fermi acceleration (see de Gouveia dal Pino \& Lazarian 2005, Lazarian 2005).

In what follows we present the observational evidence for the existence of the cosmic ray excess in the direction of the solar system magnetotail in \S \ref{sec:obs}, discuss existing explanations of this excess in \S \ref{sec:interp}. The structure of the magnatotail with magnetic field reversals arising from the solar cycle is presented in \S \ref{sec:magfi} and the mechanism of acceleration of cosmic rays in the magnetotail is outlined in \S \ref{sec:magrec}. The discussion of the results and a short summary are given by \S \ref{sec:disc} and \S \ref{sec:summ}, respectively.

\section{The observations}
\label{sec:obs}

\begin{figure}[!t]
\begin{center}
\includegraphics[width=\columnwidth]{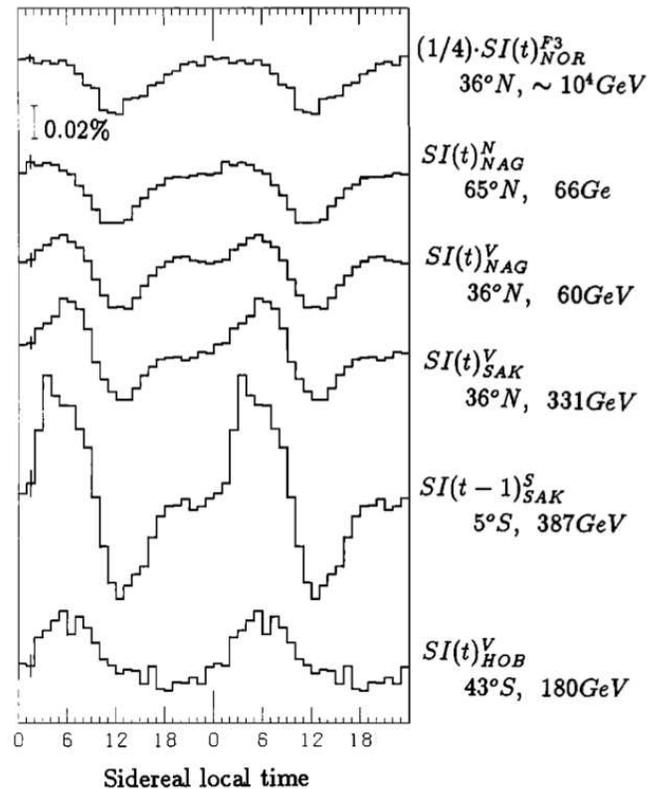}
\caption{\small Cosmic ray sidereal daily variation S(t) from different muon telescope station or air shower detectors : from top to bottom, Mount Norikura air shower array (F3, shown for reference at high energy and with the amplitude multiplied by 1/4 to account for the larger anisotropy amplitude at high energy), Nagoya station (looking 30$^{\circ}$ north-ward, and looking vertically up-ward), Sakashita station (looking vertically up-ward and $41^{\circ}$ south-ward) and Hobart station (looking vertically up-ward). Each station reports the corresponding latitude of the directional detection and the cosmic ray median energy it is sensitive to. In order to clearly show the peaks and valleys, the 24 hour variation is repeatedly  shown in a 2 days time interval. The error expresses the dispersion of the hourly relative intensity. From \citet{nagashima}}
\label{fig:nagashima}
\end{center}
\end{figure}

The observation of the large angular scale anisotropy of sub-TeV cosmic rays \citep{nagashima,hall} revealed the evidence of a superposition of two different modulations in arrival direction.
One with a sidereal variation identified with an extended deficit centered around 12 hours that seems to extend mostly across the northern hemisphere (the so-called loss cone). And one with a sidereal variation identified with a broad excess centered around 6 hours, with half opening angle of about 68$^{\circ}$ that comprises the direction of the heliotail, and extended across part of the northern and the southern hemispheres (the tail-in excess). Figure \ref{fig:nagashima} shows the combined observations of the anisotropy of sub-TeV cosmic rays from telescopes at different latitudes.

The global anisotropy amplitude is found to increase with energy up to about 5-10 TeV, however while the loss-cone structure seems to maintain a similar shape up to the multi-TeV range, the tail-in excess is still somewhat persistent in the multi-TeV range, but its broad structure appears to dissolve to smaller angular scale spots \citep{amenomori}. The apparent seasonal modulation of the tail-in excess, with a minimum amplitude in summer and a maximum (a factor of four larger) in winter, provides a compelling connection to the heliotail. 


\begin{figure}[!t]
\begin{center}
\includegraphics[width=\columnwidth]{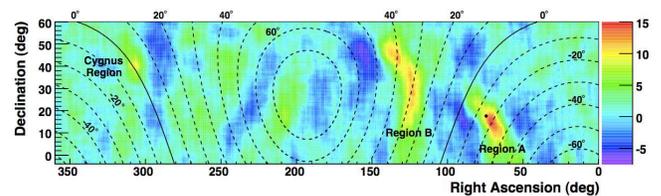}
\caption{\small Map of statistical significances from MILAGRO. A 10$^{\circ}$ bin was used to smooth the data, and the color scale gives the significance. The solid line indicates the galactic plane and the dash lines the galactic latitudes. The black dot indicates the heliotail. The fractional excess of region A is $\sim 6\times 10^{-4}$, and of region B is $\sim 4\times 10^{-4}$. From \citet{abdo2}}
\label{fig:milagromap}
\end{center}
\end{figure}

\begin{figure}[htbp]
\begin{center}
\includegraphics[scale=0.25]{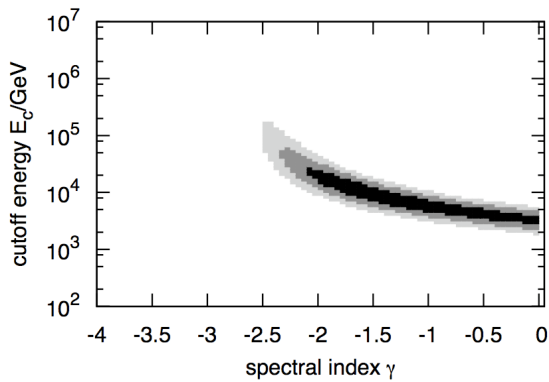}
\caption{\small Result of a $\chi^2$ fit to the excess in region A (from top of Figure 4 in \citep{abdo2}). The 1$\sigma$, 2$\sigma$ and 3$\sigma$ allowed regions of the spectral index $\gamma$ and cut-off energy $E_c$ are indicated by the shaded regions (from black to gray, respectively).}
\label{fig:regiona}
\end{center}
\end{figure}

Figure \ref{fig:milagromap} shows the multi-TeV cosmic ray arrival direction map, from the MILAGRO collaboration, obtained by eliminating anisotropies with angular structures wider than $\sim$30$^{\circ}$. The small scale structure is evidenced in this map and it shows two highly significant (more than 12 $\sigma$) localized excess regions in the cosmic rays arrival direction. Both regions are inconsistent with gamma ray emission with high confidence and therefore are claimed to be dominated by cosmic rays. They are found to have a constant yearly excess over the seven year period of collected data, however both of them were lowest in summer and highest in winter, with a $\chi^2$ probability relative to a constant fractional excess of only 5\% in each region. The strongest and more localized of them (called region A, with a fractional excess of $\sim 6\times 10^{-4}$) coincides with the direction of the heliotail (the black dot in Figure \ref{fig:milagromap}, with right ascension $\alpha \approx 74^{\circ}$ and declination $\delta \approx +17^{\circ}$ in equatorial coordinates). The corresponding energy spectrum was de-convoluted using the energy-dependent experimental observables. Figure \ref{fig:regiona} shows the result of the $\chi^2$ fit to the excess in region A assuming a pure proton spectrum of the form $E^{\gamma}\cdot e^{-{E\over E_c}}$, where $\gamma$ is the cosmic ray spectral index and $E_c$ the cut-off energy.

Figure~3 testifies that the estimated cosmic ray energy spectrum is consistent to harder than the isotropic flux (at a 4.6 $\sigma$ level) with a cutoff, and with the most significant excess in the multi-TeV range \citep{abdo2}. 

The localized excess regions of multi-TeV cosmic rays cover the same portion of the sky where the tail-in excess was observed at cosmic ray energies below TeV, and have similar seasonal modulations. It is likely that these are two manifestations of the same phenomenology and that the heliotail has an important role.

IceCube has recently reported a first view of the multi-TeV medium scale anisotropy in the southern hemisphere (i.e. only modulations that are smaller than 60$^{\circ}$) \citep{beyond2010}, which might add novel information to this observation.

\section{Other interpretations}
\label{sec:interp}

While no explanation has ever being attempted to explain the broad sub-TeV tail-in excess, a number of interpretations have been provided to address the existence of the most significant localized excess of multi-TeV cosmic rays.

Some proposed models rely on astrophysical origin. In \citet{salvati} it is noted that the two localized excess regions observed by MILAGRO surround the present day apparent location of Geminga pulsar. The supernova that gave birth to the pulsar exploded about 340,000 years ago, when its distance to the Sun was estimated to be about 90 pc. Even if the proper motion of the pulsar induced by the explosion moved it further away (the present distance of Geminga pulsar is estimated to be about 155$\pm$35 pc), 10 TeV cosmic rays produced by the supernova have propagated about 65 pc away, if we assume Bohm diffusion\footnote{Bohm diffusion assumes scattering of a cosmic ray for every gyration (see Parker 1979).} from the source to here : approximately consistent with the distance of Geminga at the time of explosion. From this distance a total cosmic ray energy of about $1.5\cdot 10^{49}$ erg must be emitted by the supernova to produce the observed fractional excess. This value is consistent with the commonly required efficiency ($\sim$ 1\%) with which a supernova energy output must be converted into cosmic rays if they are to maintain the galactic cosmic ray density.

The major problem with this explanation is that Bohm diffusion through such large distances cannot possibly explain the localized nature of the observed excesses, but it would rather produce at most a broad faint dipolar anisotropy in arrival direction. In addition, if Bohm scaling for the scattering in the immediate vicinity of the supernova shock is plausible, it seems unlikely that this regime persists during the propagation of cosmic rays through the interstellar medium \citep{drury}. On the other hand the structure of the interstellar magnetic field would very hardly maintain multi-TeV cosmic rays focussed within a 10$^{\circ}$ beam. The supposed opposite scenario of a free-streaming of cosmic rays along a sort of magnetic "nozzle" \citep{drury}, that would explain the localized nature of the observation, would also be extremely unlikely also because the propagation would have been so fast that we would not have the observation in the first place anymore. It is possible to argue a scenario with a combination of slow Bohm diffusion regime (close to the supernova due to turbulence induced by the explosion) and fast free-streaming along magnetic field through the interstellar medium (which could partially explain the localized natures of the observed excesses). But this interpretation would have problems, as the time of propagation from the source should be made long, which contradicts the idea of localization of the intensive scattering only near the source. Perhaps some sort of leaky magnetic field bottles are formed near the source, which make the propagation slow compared to the Bohm diffusion time, thus mitigating the problem.
However this possibility would require fine tuning and we have not seen this idea discussed in the literature. 

The coincidence of the most significant localized excess observed by MILAGRO with the heliotail, supports the idea that the heliosphere could somehow have a role. The possibility that we are seeing the effects of neutron production in the gravitationally focussed tail of the interstellar material was considered by \citet{drury}. As the Solar system surrounded by Solar Wind moves through the interstellar medium, the complex interaction between the two media create the heliotail. Cosmic rays propagating through the direction of the tail interact with the matter and magnetic fields to produce neutrons and hence a localized excess of cosmic ray in that direction. But while the target size has about the right size compared to the decay length of multi-TeV neutrons ($\sim$ 0.1 pc), the increase of the gravitating matter density is too low to account for the observed excess.

While it is possible to argue that the large angular scale anisotropy of cosmic rays arrival direction might be generated by a combination of astrophysical phenomena, such as the distribution of nearby recent supernova explosions \citep{erlykin}, propagation effects (Battaner, Castellano \& Masip 2009, Malkov et al. 2010) and the structure of the interstellar magnetic field, it is more likely that small angular scale anisotropies are generated by some local effect.

\section{Magnetic field structure at the helio-tail}
\label{sec:magfi}

\begin{figure}[!t]
\includegraphics[width=\columnwidth]{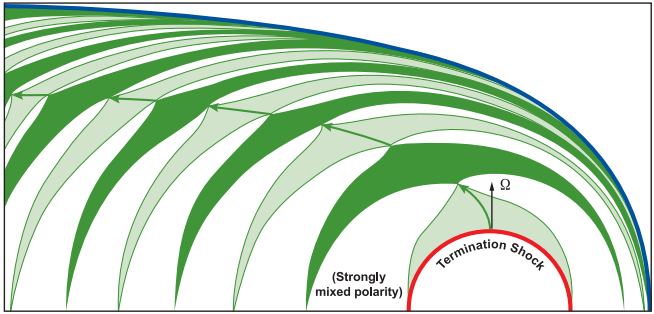}
\caption{ A meridional view of the boundary sectors of the 
heliospheric currenty sheet  and how the opposite sectors get tighter 
closer to the heliopause and into the heliotail. 
The thickness of the outflow regions in the 
reconnection region depends on the level of turbulence. The length of 
the outflow regions $L$ depends on the mean geometry of magnetic 
field and turbulence Adapted from  Nerney et al. 1995 and also Lazarian \& Opher 2009.}
\label{structure1}
\end{figure}

Figure~\ref{structure1} represents the possible structure of the heliotail which arises from the solar magnetic field cycles (Parker 1979). The magnetic fields of the opposite polarities emerge as the result of 11 year solar dynamo cycle. As the magnetic field is carried away by solar wind, the reversed magnetic field regions get accumulated in the magnetotail region. This is where reconnection is expected to occur.

Naturally, the actual heliotail is going to be turbulent, which is not represented by the idealized drawing in Figure~\ref{structure1}. As the Alfven speed is smaller than the Solar wind speed, magnetic reconnection does not change the overall magnetic field structure. Nevertheless, as we discuss in \S 5, the effects of turbulence are very important from the point of view of magnetic reconnection and the particle acceleration that it entails. 

The simulations of the magnetotail are extremely challenging (see Pogorelov et al. 2009ab) and have not been done with the sufficient resolution and extent. While we believe that future research will provide details necessary for quantitative modeling, the schematic representation of the magnetotail structure depicted in Figure~\ref{structure1} is true in terms of major features. In what follows, it will be used for describing the scenario for the origin of the cosmic ray excess that we advocate in this paper.

\section{Magnetic reconnection and cosmic ray acceleration}
\label{sec:magrec}

Astrophysical plasmas are often highly ionized and highly magnetized (Parker 1970).  The evolution of the magnetic field in a highly conducting fluid can be described by a simple version of the induction equation
\begin{equation}
\frac{\partial \vec{B}}{\partial t} = \nabla \times \left( \vec{v} \times \vec{B} - \eta \nabla \times \vec{B} \right) ,
\end{equation}
where $\vec{B}$ is the magnetic field, $\vec{v}$ is the velocity field, and $\eta$ is the resistivity coefficient.  Under most circumstances this is adequate for discussing the evolution of magnetic field in an astrophysical plasma.  When the dissipative term on the right hand side is small, as is implied by simple dimensional estimates, the magnetic flux through any fluid element is constant in time and the field topology is an invariant of motion.   On the other hand, reconnection is observed in the solar corona and chromosphere (Innes et al. 1997, Yokoyama \& Shibata 1995, Masuda et al. 1994, Ciaravella \& Raymond 2008), its presence is required to explain dynamo action in stars and galactic disks (Parker 1970, 1993), and the violent relaxation of magnetic fields following a change in topology is a promising process for the First order Fermi acceleration of high energy particles  in the universe (de Gouveia Dal Pino \& Lazarian 2003, henceforth GL03, 2005, Lazarian 2005, Drake et al. 2006, Lazarian \& Opher 2009, Drake et al. 2010). Quantitative general estimates for the speed of reconnection start with two adjacent volumes with different large scale magnetic field directions (Sweet 1958, Parker 1957).

\begin{figure}[!t]
\includegraphics[width=\columnwidth]{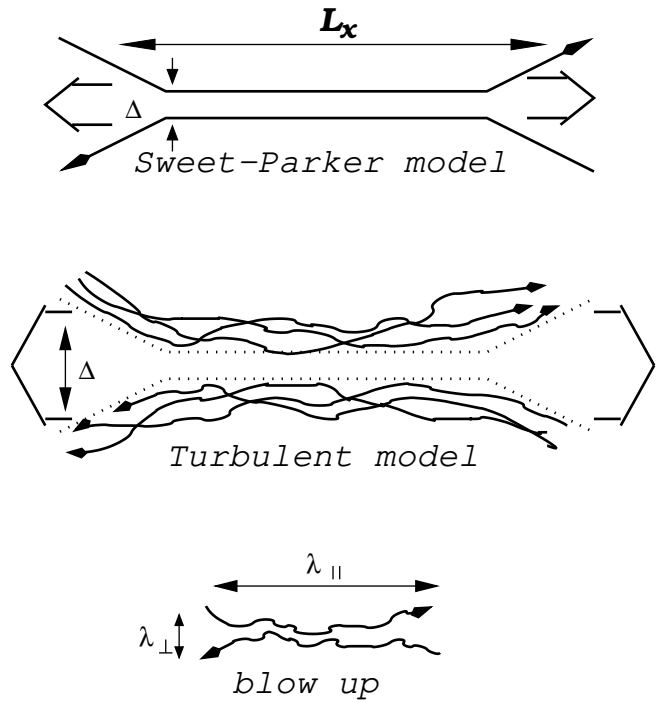}
\caption{{\it Upper plot}: 
Sweet-Parker model of reconnection. The outflow
is limited by a thin slot $\Delta$, which is determined by Ohmic 
diffusivity. The other scale is an astrophysical scale $L\gg \Delta$.
{\it Middle plot}: Reconnection of weakly stochastic magnetic field according to 
LV99. The model that accounts for the stochasticity
of magnetic field lines. The outflow is limited by the diffusion of
magnetic field lines, which depends on field line stochasticity.
{\it Low plot}: An individual small scale reconnection region. The
reconnection over small patches of magnetic field determines the local
reconnection rate. The global reconnection rate is substantially larger
as many independent patches come together. The bottleneck for the process is given by
magnetic field wandering and it gets comparable to  $L$ as the turbulence injection velocity 
approaches the Alfvenic one. From Lazarian, Vishniac \& Cho (2004).}
\label{fig:recon1}
\end{figure}

The speed of reconnection, i.e. the speed at which inflowing magnetic field is annihilated by Ohmic dissipation, is roughly $\eta/\Delta$, where $\Delta$ is the width of the transition zone (see Figure~\ref{fig:recon1}).  Since the entrained plasma follows the local field lines, and exits through the edges of the current sheet at roughly the Alfven speed, $V_A$, the resulting reconnection speed is reduced compared to the Alfven speed by a large factor. This factor in the textbook Sweet-Parker model of reconnection is $S^{1/2}$, where $S\equiv (LV_A/\eta)$ is the Lunquist number, where $L$ is the length of the current sheet (Sweet 1958, Parker 1957, see also Parker 1979).  

In general, satisfying the conservation of mass condition dictates that $V_{rec}\sim V_A (L/\Delta)$. Observations require a speed close to $V_A$, so this expression implies that $L\sim \Delta$, i.e. that the region of over which magnetic flux tubes intersect should be comparable with the outflow region. This can be achieved either via making $L$ as small as the Ohmic diffusion region i.e. that the magnetic field lines reconnect in an ``X point'' (Petscheck 1964), or the outflow region should get increased dramatically beyond the size that is predicted in the Sweet-Parker model. While for years the problem of fast reconnection was viewed as connected with proving of the stability of X point, the situation has changed recently with the second way of dramatically increasing the thickness of the outflow is becoming more popular and getting observational support (see Ciaravella \& Raymond 2008).   

The first model of X point reconnection was proposed by Petschek (1964).  In this case the reconnection speed may have little or no dependence on the resistivity\footnote{In general, the reconnection is termed fast when the reconnection velocity does not depend on the Lundquist number $S$ or if it depends on $\ln(S)$. In all other cases the large values of $S$ make reconnection too slow for most of astrophysical applications.}.  The X point configuration is known to be unstable to collapse into a sheet in the MHD regime (see Biskamp 1996), but in a collisionless plasma it can be maintained through coupling to a dispersive plasma mode (Sturrock 1966). Recent years, have been marked by the progress in understanding some of the key processes of reconnection in astrophysical plasmas. In particular, a substantial progress has been obtained by considering reconnection in the presence of Hall effect, which is described by the ${\bf J}\times{\bf B}$ term in Ohm's law:
\begin{equation}
{\bf E}+\frac{{\bf v}\times{\bf B}}{c}-\frac{{\bf J}\times {\bf B}}{e n_e c}=\frac{4\pi\eta {\bf J}}{c^2}
\end{equation}
where $e$ is electron charge and $n_e$ is concentration of electrons. Numerical experiments showed that Hall-MHD reconnection is capable of supporting X-points and thus can make the reconnection fast, i.e. comparable to the Alfven speed (Shay et al. 1998, 2004).

The condition at which Hall-MHD term gets important for the reconnection is that the ion skin depth $\delta_{ion}$ is comparable with the Sweet-Parker diffusion scale $\Delta$. The ion skin depth is a microscopic characteristic and it can be viewed at the gyroradius of an ion moving at the Alfven speed, i.e. $\delta_{ion}=V_A/\omega_{ci}$, where $\omega_{ci}$ is the cyclotron frequency of an ion. In the heliotail for a proton we find that $\delta_{ion}\sim 10^3$~km. Thus one can get the constraint on the scale $L$ for which Hall-MHD effects should dominate the reconnection:
\begin{equation}
\frac{\Delta}{\delta_{ion}}\approx 0.2 \left(\frac{L}{\lambda_{mfp}}\right)^{1/2}\beta_{pl}^{1/4}<1,
\label{constraint}
\end{equation}
where $\lambda_{mfp}$ is the electron mean free path, where $\beta_{pl}$ is the ratio of thermal pressure to magnetic pressure (see more discussion in Yamada et al. 2006). We agrue in
Appendix A that in realistic situations in turbulent media the scales $\lambda_{\|}=L_{turb}$ (see Figure~\ref{fig:recon1}) over which the microscale Sweet-Parker reconnection of individual turbulent patches may take place are much smaller than the scale of the system and therefore the collisionless effects take place within the heliotail. This, as we argue below, does not change the overall rates of magnetic reconnection.  

A shortcoming of many discussions of magnetic reconnection is that the traditional setup does not include ubiquitous pre-existing astrophysical turbulence\footnote{The set ups where instabilities play important role include Simizu et al. (2009a,b). For sufficiently large resolution of simulations those set-ups are expected to demonstrate turbulence. Turbulence initiation is also expected in the presence of plasmoid ejection (Shibata \& Tanuma 2001). Numerical viscosity constrains our ability to sustain turbulence via reconnection, however.} (see Armstrong, Rickett \& Spangler 1994, Elmegreen \& Scalo 2004, McKee \& Ostriker 2007, Lazarian 2009, Chepurnov \& Lazarian 2010). As turbulence radically changes many astrophysical processes, the influence of turbulence on reconnection has attracted the attention of researchers for a long time (see Speizer 1970, Straus 1988). An extended discussion of turbulence role in accelerating reconnection can be found in Mathaeus \& Lamkin (1985, 1986). However, there the X-points created by turbulence together with the effects of compressibility and heating were identified as the means of accelerating reconnection.

A very different approach to the effects of turbulence was adopted in Lazarian \& Vishniac (1999, henceforth LV99). Their model does not appeal to any of the effects of turbulence-created X-points, compressibility or heating and it is applicable to a wide range of astrophysical conditions. Fortunately, this approach provides a robust way of accelerating reconnection. Indeed, as we mentioned above, the approach in LV99 is to consider ways to decouple the width of the plasma outflow region from the scale determined by Ohmic effects\footnote{In the Sweet-Parker reconnection both widths coincide and equal to $\Delta$. In the LV99 model the outflow $\Delta$ is much wider than the thickness of the individual current sheets, while the Sweet-Parker is obtained as a degenerate case of no turbulence.} The plasma is constrained to move along magnetic field lines, but not necessarily in the direction of the mean magnetic field.  In a turbulent medium the two are decoupled, and fluid elements that have some small initial separation will be separated by a large eddy scale or more after moving the length of the current sheet.  As long as this separation is larger than the width of the current sheet, the result will not depend on $\eta$. The mental picture presented in LV99 is that the fluid follows magnetic field lines, which are not straight, but wander (see Figure~\ref{fig:recon1} and also Lazarian et al. 2004, where this wandering was calculated numerically and compared with the analytical predictions in LV99). As a result, the thickness $\Delta$ of the fluid outflow is determined not by microphysical Ohmic diffusivity, but magnetic field wandering which for the injection velocity $V_l$ of the order of $V_A$ is of the order of the turbulence injection scale $l$, i.e. $\Delta \approx l$. If the length of the current sheet $L$ is of the order of $l$ it is clear from Figure~\ref{fig:recon1} that $V_{rec}$ can be comparable with $V_A$. Note that LV99 considers generic 3D configurations of magnetic fluxes with non-zero magnetic guide, i.e. shared, magnetic field. The shared magnetic field is being ejected from the reconnection region together with plasmas. The cosmic rays that we consider in this paper stay entrained on magnetic field lines. 

Two effects are the most important for understanding of the nature of reconnection in LV99. First of all, in three dimensions bundles of magnetic field lines can enter the reconnection region and reconnect there independently (see Figure~\ref{fig:recon1}), which is in contrast to two dimensional picture where in Sweet-Parker reconnection the process is artificially constrained. Then, the nature of magnetic field stochasticity and therefore magnetic field wandering (which determines the outflow thickness, as illustrated in Figure~\ref{fig:recon1}) is very different in 2D and the real 3D world (LV99). In other words, by removing artificial constraints on the dimensionality of the reconnection region and the magnetic field being absolutely straight, LV99 explore the real-world astrophysical reconnection, which has become a topic of modern numerical studies\footnote{The issue of what is happening in 2D reconnection in presence of turbulence is rather controversial (see Loreiro et al. 2009, Kulpa-Dubel et al. 2010). Unlike 3D reconnection, the X-points must play role there if the reconnection is fast (Matthaeus \& Lamkin 1986, Servidio et al. 2009). However as the nature of turbulence and reconnection are different in 2D, we feel that the 2D studies cannot clarify much in the physics of the actual astrophysical reconnection.} (see Dorelli \& Bhattacharjee 2008, Kowal et al. 2009, Daughton et al. 2009).

Analytical calculations in LV99 showed that the resulting reconnection rate is limited only by the width of the outflow region.  That model predicts reconnection speeds close to the turbulent velocity in the fluid.  More precisely, assuming isotropically driven turbulence characterized by an injection scale, $l$, smaller than the current sheet length $L$, LV99 obtained
\begin{equation}
V_{rec}\approx V_A\left(l/L\right)^{1/2}\left(V_l/V_A\right)^2,
\label{eq:recon0}
\end{equation}
where the turbulent injection velocity $V_l$ is assumed to be less than $V_A$. If $L<l$, the first factor in Eq.~(\ref{eq:recon0}) should be changed to 
$(L/l)^{1/2}$ (LV99). Taking into account that $V_{turb}=V_A (V_l/V_A)^2$ is the velocity at the scale where turbulence transits from weak MHD turbulence to the strong MHD turbulence\footnote{Weak and strong do not reflect the amplitudes of waves, but the interaction between the oppositely moving Alfvenic wave packages. For our case the largest amplitude waves are in the regime of weak turbulence, but smaller amplitude small scales ripples are in the regime of strong turbulence.} (see LV99, Lazarian 2006), one can rewrite Eq.~(\ref{eq:recon0}) in the following way: 
\begin{equation}
V_{rec}\approx V_{turb}\left(l/L\right)^{1/2}, 
\label{eq:recon1}
\end{equation}
Note, that here "strong" means only that the eddies decay through nonlinear interactions in an eddy turn over time (see more discussion of the  LV99).  All the motions are weak in the sense that the magnetic field lines are only weakly perturbed. The predictions of the LV99 model including the analytical scaling given by Eq.~\ref{eq:recon1} have been successfully tested\footnote{Testing of whether reconnection is fast, i.e. independent of resistivity, is tricky with present day diffusive codes. Therefore it is the successful testing of the analytical predictions that give us confidence in the results.} with extensive MHD simulations in Kowal et al. (2009). 

It is clear from Eq.~(\ref{eq:recon1}) that the level of turbulence controls the reconnection rate in the LV99 model. If the turbulence level is low and  the region of magnetic field diffusion, i.e. field wandering region, is thinner than the $\Delta$ given by Ohmic diffusion, the reconnection happens with the Sweet-Parker rate. This may explain the accumulation of magnetic flux prior to Solar flares. However, outflows can destabilize the system by inducing turbulence and increasing reconnection/outflow rate (see Lazarian \& Vishniac 2009).  We, however, do not believe that the case of marginal initial turbulence is applicable to the magnetotail. 

The LV99 model is a model of volume-filled reconnection with magnetic filaments/flux tubes filling the wide outflow region\footnote{We would like to stress that Figure~\ref{fig:recon1} exemplifies only the first moment of reconnection when the fluxes are just brought together. As the reconnection develops the volume of thickness $\Delta$ gets filled with the reconnected 3D flux ropes moving in the opposite directions.}. The LV99 presented such a model of reconnection and observations of the Solar magnetic field reconnection support the volume-filled idea (Ciaravella \& Raimond 2008). In this wide region magnetic field lines shrink, converting magnetic energy into other forms. In the absence of cosmic rays, the magnetic energy is being transfered into kinetic energy of the outflow, but in the presence of energetic particles a portion of this energy can be channeled into the acceleration of these particles.

\begin{figure}[!t]
\begin{center}
\includegraphics[width=0.8\columnwidth]{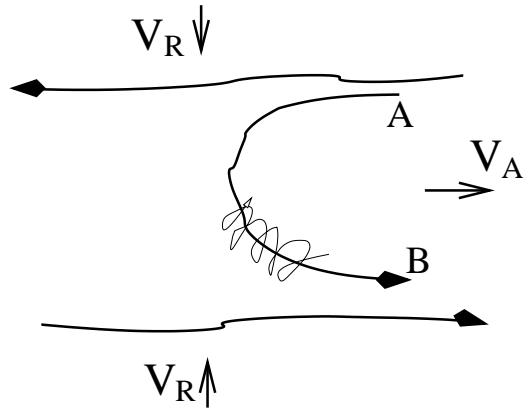}
\caption{  
Cosmic rays spiral about a reconnected magnetic
field line and bounce back at points A and B. The reconnected
regions move towards each other with the reconnection velocity
$V_R$. The advection of cosmic rays entrained on magnetic field
lines happens at the outflow velocity, which is in most cases
of the order of $V_A$. Bouncing at points A and B happens
because either of streaming instability induced by energetic particles or magnetic
turbulence in the
reconnection region. In reality, the outflow region gets filled in by the oppositely moving tubes of reconnected flux which collide only to repeat on a smaller scale the pattern of the larger scale reconnection. Thus our cartoon also illustrates the particle acceleration taking place at smaller scales. From Lazarian (2005).}
\label{recon2}
\end{center}
\end{figure}

Figure~\ref{recon2} exemplifies the simplest realization of the acceleration within the reconnection region expected within LV99 model. As a particle bounces back and forth between converging magnetic fluxes, it gains energy through the first order Fermi acceleration described in de Gouveia dal Pino \& Lazarian (2003, 2005, henceforth GL03\footnote{The mechanism published eventually in 2005 paper was made public as 2003 arXiv preprint, which motivates us
to refer to the original on-line publication.}, see also Lazarian 2005). The first order acceleration of particles entrained on contracting magnetic loop can be understood from the Liouville theorem, i.e the preservation of the phase volume which includes the spatial and momentum coordinates. As in the process of reconnection the magnetic tubes are contracting and the configuration space presented by magnetic field shrinks, the regular increase of the particle's energies is expected. The requirement for the process to proceed efficiently is to keep the accelerated particles within the contracting magnetic loop. This introduces limitations on the particle diffusivities perpendicular to magnetic field direction.   
The subtlety of the point above is related to the fact that while in the first order Fermi acceleration in shocks magnetic compression is important, the acceleration via LV99 reconnection process is applicable to incompressible fluids. Thus, unlike shocks, it is not the entire volume that shrinks for the acceleration, but only the volume of the magnetic flux tube. Thus high perpendicular diffusion of particles may decouple them from the magnetic field. Indeed, it is easy to see that while the particles within a magnetic flux rope depicted in Figure~6 bounce back and forth between the converging mirrors and get accelerated, if these particles leave the flux rope fast, they may start bouncing between the magnetic fields of different flux ropes which may sometimes decrease their energy. Thus it is important that the particle diffusion parallel and perpendicular magnetic field stays different. Particle anisotropy which arises from particle preferentially getting acceleration in terms of the parallel momentum may also be important.  

The energy spectrum was derived in GL03 following the routine way of dealing with the first order Fermi acceleration in shocks (see Longair 1992), namely, by 
considering the acceleration rate and the loss rate of energetic particles without taking into account the backreaction of the accelerated particles on the flow. This way GL03 obtained:
\begin{equation}
N(E)dE=C\cdot E^{-5/2}dE,
\label{-5/2}
\end{equation}
i.e. the spectrum of energetic particles with the spectral index similar to that of galactic cosmic rays (see also its derivation in LO09). 

More recently, Drake, Swisdak \& Shay (2006, henceforth DSS06) approached the problem of back-reaction of accelerated particles on magnetic loops produced by reconnection\footnote{Unlike GL03 the study in DSS06 treated particle acceleration in 2D reconnection layers, where shrinking filaments degenerate into closed loops. The latter study also appeals to the physics of collisionless reconnection. However, from the point of view of the acceleration process, these differences do not matter much, provided that the reconnection processes are fast and the reconnected magnetic fields fill the outflow region. The latter condition we feel is somewhat more difficult to realize in 2D picture of DSS06, but we do not dwell on these details.}. The authors conjectured that the backreaction can be described by the term $(1-8\pi\epsilon/B^2)$, where $\epsilon$ is the energy of accelerated particles, and obained the spectrum of accelerated particles $-3/2$ rather than $-5/2$. We feel that the introduction of the backreaction into the acceleration process is an important process that requires further studies.

The idea of the acceleration of protons by magnetic reconnection has been supported by numerical simulations. In Drake et al. (2010) two dimensional calculations of the particle acceleration within contracting magnetic loops were presented. The first realistic three dimensional calculations which use the actual reconnection simulations, rather than artificially creating loops, were presented in Lazarian et al. (2010). More numerical studies of the acceleration are necessary, as the realistic simulations are extremely challenging.   

We note that the acceleration described in GL03 is different from electric field acceleration described e.g. in Haswell et al. (1992) and Litvinenko (1996). The acceleration discussed there within Sweet-Parker reconnection layers is inefficient, first of all, because only a tiny rate of magnetic energy release in reconnection processes is dictated by exceedingly slow reconnection rates. Moreover, it is rather problematic to confine the particles within the reconnection Sweet-Parker layer. Any field wiggles would naturally result in particles leaving the layer\footnote{It is shown in LV99 that the probability of magnetic field return into a reconnection layer is low. Thus the accelerated particles are bound to leave the zone of accelerating magnetic field fast without getting much of energy gain.}. As a result the acceleration gets inefficient.  

The estimate of the maximal energy of protons which can be accelerated through this process can be obtained through the usual arguments that the Larmor radius should not be larger than the size of the magnetized region $L_{zone}$ (Longair 1997)
\begin{equation}
E_{max}\approx 10^{13}~{\rm eV} \left(\frac{B}{1~{\rm \mu G}}\right) \left(\frac{L_{zone}}{2\times 10^{15}~cm}\right),
\end{equation} 
which sets the limit of energies, i.e. $E_{max}$, through appealing to the fact that protons of larger energies cannot be confined by magnetic fields to experience the acceleration through multiple bouncing back and forth between the reconnecting magnetic fluxes\footnote{This does not preclude cosmic rays of higher energies to experience additional acceleration via electric field as in Litvinenko (1996), but we do not discuss this possibility in the paper.}. 
 
The properties of the heliotail, especially at large distances from the Sun, are not well constrained at the moment (see Pogorelov et al. 2009b, Pogorelov 2010, private communication), which makes it challenging to identify the precise values for magnetic field strength and scales involved. A simple estimate of the scale $L_{zone}$ and magnetic field $B$ can be obtain if we accept the solar wind velocity 450 km/s (see Pogorelov et al. 2009b) the 11 year cycle of magnetic field can create create areas of field of one direction of the order of $10^{16}$~cm. Assuming that the magnetic field is about 1 $\mu$G, one can get particles with the energy of dozens of TeVs. While the detailed calculations, should produce more accurate estimates of the magnetic field and the scales involved, we feel that we are getting right ballpark numbers. In fact, we can predict that, unless some processes of field amplification operate in the turbulent heliotail, the acceleration of the cosmic rays of energies much larger than $10$~Tev is rather unlikely by magnetic reconnection.

\section{Discussion}
\label{sec:disc}

This paper attempts to explain the cosmic ray excess in the range from ~50 GeV to 1-10 TeV as arising from magnetic reconnection in the magnetotail. The high energy cut off observed corresponds roughly to what is expected from the reconnection events. Indeed, it is virtually impossible to explain the acceleration of higher energy particles with the mechanism, unless appealing to some hypothetical magnetic field acceleration processes. The difference in the distribution of the excess of low and higher energy particles in our scenario arises from higher efficiencies of scattering for low energy particles. 

Our relation of the observed cosmic ray anisotropies to the heliotail is also supported by the sideral daily variations of galactic cosmic rays observed with the Tsumeb neutron monitor's hourly count during 1977-2000 reported in Karapetyan (2010). There it was argued that the observed cosmic ray excess has heliotail rather than galactic origin. No physical explanation of the excess was provided there, however. On the contrary, we relate the excess of the cosmic rays with the acceleration process induced by magnetic reconnection.

The paper has an exploratory character, as the quantitative description of mechanisms of cosmic ray acceleration in the reconnection regions are at its infancy. Unlike shock acceleration, which is the subject of long history (Axford et al. 1977, Krymsky 1977, Bell 1978, Blandford \& Ostriker 1978) and extensive literature (see Gaisser 1990, Malkov \& Diamond 2009). The existing analytical models (de Gouveia dal Pino \& Lazarian 2003, Drake et al. 2006, Drake et al. 2010) are insufficiently elaborated, while the numerical simulations (see Drake et al. 2010, Lazarian et al. 2010) are rather idealized. Nevertherless, the accumulated evidence is suggestive that the process of the acceleration can be efficient.

We note that dealing with the acceleration of protons we do not distinguish between the collisionless reconnection and the LV99 model of reconnection. We feel that, provided that the reconnection regions are thick and filled with the reconnecting shrinking loops, the acceleration happens the same way for the two cases. We, however, claim that the formation of such regions may be somewhat problematic within the paradigm of collisionless reconnection. At the same time, this situation is a clear consequence of the LV99 model, which appeals to the ubiquitous astrophysical turbulence to enpower it.

While we accept that the model of acceleration in the reconnection regions require more study, we would like to stress that the proposed scenario has several attractive features. First of all, it allows us to address the issues of cosmic ray excess over the entire range of energies that these particles were observed. Moreover, the fact that the excess of the particles is observed in the direction of the magnetotail is suggestive that the processes in magnetotail are involved. In addition, the alternative mechanisms of producing the excess apparently have self-evident problems, as we discussed in the paper. We would like to stress, that none of the alternative mechanisms provides a unifying explanation for the existence of the excess over a range of energies reported by different groups (see \S 2). In this situation we think that the proposed mechanism should be considered seriously.


We argue that the localized excesses of cosmic rays in the multi-TeV range and the tail-in excess below the TeV range are related by the same phenomenology. Within our approach the localized regions of the TeV cosmic rays are related to the sites of acceleration via reconnection. The lower energy particles can be accelerated over extended regions of the magnetotail; they are also expected to experience more scattering prior to reaching the observer at the Earth. More elaborate modeling of both cosmic ray acceleration in reconnection regions and the propagation of cosmic rays in magnetotail should provide detailed predictions to be compared with observations. 

Our recent approaches to this problem combine the advances in understanding of the statistical structure of MHD turbulence, in particular, tensorial structure of Alfvenic, slow and fast modes (see Cho, Lazarian \& Vishniac 2002, Cho \& Lazarian 2002, 2003, Kowal \& Lazarian 2010), analytical description of the propagation of cosmic rays (see Yan \& Lazarian 2002, 2004, 2008), and testing of analytical predictions numerically using magnetic fields obtained through numerical simulations (see Beresnyak, Yan \& Lazarian 2010). As we described above, simulations of particle acceleration in turbulent reconnection is under way. Thus we hope to have in future a self-consistent numerically-tested picture of the acceleration and propagation. Combined with the advances in simulations of magnetotail magnetic fields (see Pogorelov et al. 2010) this gives hope that we will have a set of quantitative detailed predictions for the cosmic ray excess arising from the mechanism described in the paper.     

Numerical modeling will also clarify the role of the second order Fermi acceleration which arise from turbulence induced both by magnetic reconnection and existing within the magnetotail. The corresponding processes have been discussed at length in the literature (see La Rosa et al. 2006, Petrosian et al. 2006, Yan et al. 2008), but we expect the second order Fermi process to be less efficient than the first order Fermi acceleration and therefore to be subdominant. 

In situ measurement of the excess of the energetic particle acceleration within reconnection regions could be beneficial as well. So far, the attempts of measuring of such an excess were not successful (Gosling et al. 2005ab, 2007, Phan et al. 2007). We believe that this is due to the X-point, Petscheck-type reconnection is inefficient in the acceleration of cosmic rays\footnote{The inefficiency of X-point reconnection arises from both tiny amount of magnetic energy being released in the small reconnection region and in inefficiency of slow shocks (see Beresnyak, Jones \& Lazarian (2010) in accelerating energetic particles. An alternative explanation of the fact was given by H. Karimabadi (private communication) who observed in his kinetic simulations extended periods of the stagnation of the X-point reconnection.} (see LO09). This type of reconnection was sought in the studies above. The mechanism we discussed above (GL03, DX06) appeals to the thick extended reconnection regions. Such regions naturally emerge in the magnetotail as turbulent magnetic fields of opposite polarity are being pressed together (see Figure~4).

The GL03 acceleration process has been already employed in Lazarian \& Opher (2009, henceforth LO09) to explain the origin of anomalous cosmic rays, whose measurements by Voyagers seem to contradict to their origin within most of the accepted models of shock acceleration. A similar conclusion was also obtained in Drake et al. (2010) where the process of collisionless, but volume-filling\footnote{We stress this, as the original models of collisionless reconnection (see Shay et al. 1998, 2004), unlike LV99, stressed the importance of pointwise, i.e. X-point, reconnection.} reconnection was discussed. Whether small-scale reconnection is collisionless or collisional does not play a role for LV99 model. The testing of this fact was successfully performed in Kowal et al. (2009), where the plasma effects were simulated through the introduction of the anomalous resistivity. 

Unlike LO09, where the acceleration of energetic particles in the reversing field of the heliosheath was considered, we consider the magnetic field reversals in the heliotail. While the field reversals in the heliosheath arise from Sun's rotation with the magnetic axis being not parallel to the rotation axis, the reversals in the heliotail arise from the 11 year solar cycles. As a result, the scale of the reversals is expected to be much larger. This provides a possibility of accelerating higher energy particles.      

\section{Summary}
\label{sec:summ}

In this paper we combine data from different experiments to prove that there exists a statistically significant excess of energetic particles in the direction of the solar system magnetotail. We proposed an explanation of this excess as arising from the acceleration of energetic particles in reconnection regions along the magnetotail. These regions arise as oppositely directed magnetic field of the solar wind is pressed together in the magnetotail. The change of the magnetic field polarity arises in our scenario due to the well established solar cycles.

\acknowledgments

We thank our colleagues in Ice Cube collaboration, in particular, Francis Halzen, for numerous fruitful discussions. AL acknowledges the support of the NSF grant AST 0808118, NASA grant X5166204101 and of the NSF-sponsored Center for Magnetic Self-Organization. PD acknowledges the support from the U.S. National Science Foundation-Office of Polar Programs.

\appendix
\section{Collisionless effects in magnetotail reconnection}

While the LV99 model provides fast reconnection, i.e. the reconnection that does not depend on resistivity, without appealing to any collisionless plasma effects, for some of the small-scale events, e.g. for the acceleration of low energy electrons the microphysics of reconnection may be important. In the reconnection process described by LV99 model the reconnection speed is given by Eq.~(\ref{eq:recon1}). The same speed can be obtained if one considers the local reconnection of flux tubes. It was shown in LV99 that the probability of magnetic field lines which ones entered the reconnection layer to reenter the reconnection layer again is low. As a result it was shown in LV99 that the global reconnection rate is
\begin{equation}
V_{rec, global}\approx L/\lambda_{\|} V_{rec, local}
\label{global}
\end{equation} 
where $V_{rec, local}$ is the velocity of reconnection within the small-scale local Sweet-Parker reconnection events, i.e. reconnection events on scale $\lambda_{\|}$ depicted on the lower panel of Figure \ref{fig:recon1}. It is easy to show that assuming that if $\lambda_{\|}$ and $\lambda_{\bot}$ are related through the GS95 critical balance, namely $\lambda_{\|}/V_A\approx \lambda_{\bot}/v$, Eq.~(\ref{global}) results in too high reconnection rates even for Sweet-Parker reconnection at 
scales $\lambda_{\bot}$. The naively obtained rates much exceed those provided by Eq.~(\ref{eq:recon1}), which proves that the Ohmic resistivity effects are not the bottleneck of the LV99 model (see more details in LV99). 

We will use Eq.~(\ref{global}) (\ref{eq:recon1}) to establish the scale of the local Sweet-Parker events $\lambda_{\|}$:
\begin{equation}
\lambda_{\|}\approx LS^{-1/3}\left(L/l\right)^{1/3}\left(V_A/V_l\right)^{4/3}
\end{equation}
which results in the thickness of the Sweet-Parker layers being
\begin{equation}
\Delta_{turb}\approx LS^{-2/3} \left (L/l\right)^{1/6} \left(V_A/V_l\right)^{2/3}.
\end{equation}
Those $\Delta_{turb}$ correspond to $\lambda_{perp}$ in the lower panel of Figure~\ref{fig:recon1}.

The corresponding thickness is much smaller that the thickness of the laminar Sweet-Parker layer, which is $L S^{-1/2}$, which makes according to Eq.~(\ref{constraint}) the collisionless effects important for the small scale heliotail reconnection. This, however, is not going to change either the rate of magnetic reconnection or the acceleration of protons. The presence of collisionless effects can affect the acceleration of electrons, which is, however, not the process that we deal in this paper. 

We should add that the issue of fluid being collisionless or collisional is not so simple in the presence of turbulence. Collisionless fluids are subject to instabilities, which reduce the effective degree of their collisionality via inducing resonant scattering of particles. Such collisions mediated by magnetic field decrease the mean free path of particles. For instance, in Lazarian \& Beresnyak (2006) the problem of the collisions in a fluid of was treated self-consistently in the presence of the gyroresonance instability. It was demonstrated there that turbulent compressions of the fluid on the mean free path induce instability at the gyroradius which decrease the mean free path. For thermal plasma particles firehose and mirrow instability should also be important (see Schekochihin et al. 2010). As a result, we can state that turbulence both decreases the scale over which elementary reconnection events take place, potentially allowing the elementary reconnection events to proceed in a collisionless fashion, but at the same the compressions arising from the turbulent cascade\footnote{Slow and fast modes are mostly responsible for the compressions in sub- and trans-Alfvenic turbulence (see Cho \& Lazarian 2003).} can decrease the mean free path of the particles. The details of these interesting processes are not important for the acceleration of cosmic ray protons that we deal in this paper.

\end{document}